\documentclass[preprint,showpacs,showkeys,amsmath,amssymb]{revtex4}
\usepackage{bm}
\usepackage{amsmath}
\usepackage{amssymb}
\usepackage{graphicx}
\begin{document}

\title{Synchronized oscillations on a Kuramoto ring and their entrainment under periodic driving}
\author{Tarun Kanti Roy}
\email{tarun.roy@saha.ac.in}
\affiliation{Saha Institute of Nuclear Physics, Kolkata 700064, INDIA}
\author{Avijit Lahiri}
\email{a_l@vsnl.com}
\affiliation{Dept of Physics, Vidyasagar Evening College, Kolkata 700006, INDIA}

\begin{abstract}
\vskip .5cm
\noindent We consider a finite number of coupled oscillators as an adaptation of the Kuramoto model of populations of oscillators. The synchronized solutions are characterized by an integer $m$, the winding number, and a second integer $l$. Synchronized solutions of type ($m$, $l=0$) are all stable, and an explicit perturbative expression for these for large values of the coupling constant $K$ is presented. For low $K$, these solutions appear at certain specific values, each merging with a solution of type ($m$, $l=1$), both these solutions being stable for $K$ close to the relevant value. The ($m$, $0$) solution continues to be stable for larger $K$, while the ($m$, $1$) solution, on continuation in $K$ becomes unstable and then merges with a new type (with a different $m$ and $l$) of unstable solution. The ($m$,$0$) type solutions for large $K$ are in the nature of phase waves traveling round the ring. All the stable synchronized solutions are entrained by an external periodic driving, provided that the driving frequency is sufficiently close to the frequency of the synchronized population. A perturbative approach is outlined for the construction of the entrained solutions. For a given amplitude of driving, there is a certain maximum detuning between the two up to which the entrained solution persists. The question of stability of the entrained solution is addressed. The simplest situation involving three oscillators is investigated in details where the onset of synchronization is seen to occur through a tangent bifurcation at some critical value of $K$. Immediately before the appearance of the first synchronized solution, the system exhibits intermittent chaos with a typical scaling for the duration of the laminar phase. Under a periodic driving with an appropriately limited detuning, there occurs entrainment of the chaotic solution.
\end{abstract}

\maketitle

\newcommand{\vn}{\vskip .5cm\noindent}
\newcommand{\nn}{\noindent}
\newcommand{\bs}{\begin{subequations}}
\newcommand{\es}{\end{subequations}}
\newcommand{\fs}{\footnotesize}
\newcommand{\ns}{\normalsize}

\section{Introduction}

\vn {\it Synchronization} in coupled non-linear oscillators has been observed and investigated for a long time now. Starting from the observation by Christian Huyghens in 1665~\cite{chaos09,strogatz} of synchronization in weakly coupled pendulum clocks, a wide variety of phenomena involving synchronization has been explored, including the spectacular synchronization of the flashing of fireflies~\cite{buck}. In recent decades, a theoretical understanding of synchronization has played a definitive role in discerning and explaining the mechanism underlying {\it biological rhythms}~\cite{strogatz} of a wide variety. Thus, populations of self-oscillatory neuronal cells with mutual interactions exhibit the phenomenon of {\it phase locking}, where all the cells oscillate in phase, as in the cardiac pacemaker, studies on which were initiated in a pioneering work by Peskin~\cite{peskin}. Similar synchronization phenomena are found in the insulin-secreting cells in the pancreas, in populations of cells controlling such activities as breathing, walking, and running, and in the rhythmic activites of the intestines (~\cite{strogatz,mirollo,ermentrout3}).  

\vn A simple but highly fruitful model of an interacting population of nonlinear oscillators, proposed by Kuramoto (see, e.g., ~\cite{kuramoto-pap}), opened up a new era in the analysis of large populations of interacting nonlinear oscillators where synchronization appears as a macroscopic mode of behaviour for such a population. The onset of synchronization could be studied in the model in terms of a macroscopic order parameter relating to the population. A vast literature is now available on the Kuramoto model (which followed a class of models proposed earlier by Winfree~\cite{strogatz4,winfree}) and its variants, and on the explanation of synchronization phenomena in diverse areas in the physical and biological sciences on the basis of the model. A comprehensive review of various aspects of the Kuramoto model with applications in neural networks, Josephson junction arrays, laser arrays, charge density waves, and laser oscillators, is to be found in~\cite{kuramoto1}.

\vn Another area of study where the Kuramoto model is potentially of great use is the phenomenon of {\it entrained synchronization} where the effect of externally imposed periodic driving on a synchronized population of coupled nonlinear oscillators is explored with the help of the model. Entrainment is a term generally used to denote the influence of two or more oscillating systems on one another whereby they oscillate at a common frequency. According to this definition, the phenomenon of synchronization itself is an instance of mutual entrainment of a population of oscillators. However, we shall use the term in a more specialized sense here, where it is meant to describe the modofication of the frequency of synchronized oscillations of a population of oscillators by an externally imposed periodic influence such that the latter makes the entire population oscillate with its own frequency.

\vn This phenomenon of entrainment operates in the {\it circadian} time-keeping system of organisms where the external influence (commonly refrred to as a {\it zeitgeber} in the biological literature) is the day-night cycle of light intensity that affects the endogenous oscillations by means of signals generated through appropriate optical receptors. 

\vn The problem of exogenous entrainment of an endogenously oscillating synchronized populations of oscillators described by the Kuramoto model has been considered by Sakaguchi~\cite{sakaguchi} and more recently by Ott {\it et al}~\cite{ott5}. The organization of the circadian timing system (CTS) and a simplified model of this syetem, has been described in~\cite{circadian1}, where the pacemaker of the suprachiasmatic nucleus (SCN), a central component of the CTS, has been treated as a large but finite population of cells described by the Kuramoto model.

\vn While the approach initiated by Kuramoto and followed in a large number of subsequent investigations considers the thermodynamic limit $N\rightarrow\infty$, where $N$ stands for the number of interacting oscillators, and then looks for synchronized state in terms of a macroscopic order parameter, the behaviour of the population for a {\it finite} $N$ has also been investigated, with results analogous to the thermodynamic approach, though exact results for finite $N$ are few in number. Maistrenko {\it et al}~\cite{kuramoto5} have considered the scenario just before the onset of synchronization with global coupling (i.e., each oscillator coupled to all the remaining oscillators in the population) for finite $N$. The case $N=3$ is a well-known torus flow while $N=5$ gives important insights into the phase flow leading to synchronization. In a subsequent work, Popovych {\it et al}~\cite{kuramoto8} demonstrated the existence of a high dimensional chaotic regime in the finite-$N$ Kuramoto model, such high dimensional chaos being a common feature in coupled oscillator systems.

\vn Synchronization in finite linear chains of oscillators has been investigated in a number of papers (see~\cite{ermentrout1} and references therein) with nearest neighbour coupling, multiple neighbour coupling, and also with global coupling. Travelling wave solutions were obtained in these models, where the range of coupling was found to affect the phase lags between the successive oscillators. Such chains of oscillators are relevant as models of central pattern generators (CPG) in certain organisms (see~\cite{ermentrout2} and references therein).

\vn In the present paper we consider a finite number of coupled oscillators with a {\it ring}-like connectivity, based on a nearest-neighbour interaction, where each oscillator is represented by a time-dependent phase as in the Kuramoto model. Rings of coupled oscillators were considered in~\cite{ermentrout3}, where each oscillator was an weakly attracting limit cycle (other possible degrees of freedom having been eliminated by a centre-manifold reduction). Phase wave solutions travelling round the ring were found, and a stabilty analysis of these phase waves was also performed. Oscillating systems coupled into a ring-like topology serve as a model for a wide array of applications such as locimotion gaits, rings of semiconductor lasers, and circular antenna arrays~\cite{ring-model}. Rogge and Aeyels~\cite{rogge1} considered a chain of oscillators with {\it unidirectional} nearest neighbour couplings and derived an algorithm for the construction all the phase-locked solutions of the system. They also made a detailed stability analysis of these solutions, deriving interesting results.

\vn Our model is similar to that in~\cite{rogge1} with, however, a {\it bidirectional} coupling as in~\cite{ermentrout3}, where the system consists of $N$ oscillators as in the Kuramoto model, each represented by a time-dependent phase. The same model has been considered by Ochab {\it et al}~\cite{kuramoto6}, who derived an algorithm for the synchronized solutions of the system. We follow the approach of these authors and work out the stable solutions in the limit of large $K$, where $K$ stands for the coupling constant. For sufficiently small values of $K$, the system does not admit of synchronized solutions. At larger values of $K$, there appear synchronized solutions where each such solution is characterized by two integers $m,~l(1\leq l\leq N)$. The number and types of synchronized solutions change as $K$ is made to assume increasing values. This is illustrated in details for the simplest non-trivial case $N=3$, where a stable and unstable solution are seen to appear through a tangent bifurcation at a certain critical value of $K$ ($=K_{\rm c}$), the two solutions being characterized by the integers $m=0,l=1$. As $K$ is made to increase, the stable $(0,1)$ solution changes to a $(0,0)$ type solution at a certain vealue of $K$, while the unstable $(0,1)$ solution changes into a $(m=1,l=2)$ type solution. Similar other scenarios, for other $(m,l)$ type solutions, appear at still larger values of $K$. For $K$ slightly below $K_{\rm c}$, the system exhibits {\it intermittent chaos}, characterized by alternating laminar and turbulent phases, with a characteristic scaling of the duration time of the laminar phase and the Lyapunov exponent.

\vn An explicit perturbative construction of solutions is possible for sufficiently large values of $K$, where it is seen that only the $l=0$ type solutions (with various different values of $m$; for $N=3$, the only possible value of $m$ for $l=0$ is $m=0$) are stable. The $l=0$ solutions for various values of $m$ appear as stable phase waves traveling round the ring of oscillators. While our results constitute an extension of those in~\cite{kuramoto6}, a complete picture describing all the stable and unstable synchronized solutions for all values of $K$, and for arbitrary $N$, is yet to emerge.

\vn Finally, we extend the model to include an external periodic driving on the system and study the entrainment properties as a function of the apmlitude ($A$) and frequency ($\Omega$) of driving. For any given frequency $\Omega$, sufficiently close to the free-running synchronized frequency ($\omega$) of the system,  there exists a critical amplitude $A_0$ above which entrainment to the external frequency $\Omega$ occurs. We present a perturbative construction of the entrained solutions starting from the synchronized solutions obtained in our model. It is also found that the question of stability of the entrained solutions reduces to simple terms since stability is seen to be determined by a constant rather than a periodic matrix.

\vn The plan of the paper is as follows. In sections~\ref{formulate} and~\ref{synch} we formulate the Kuramoto ring problem consisting of $N$ oscillators with bidirectional nearest neighbour interaction, and set up the basic equation describing the synchronized solutions. Section~\ref{stability} includes the formulation of the stability problem for the possible synchronized solutions of the system. An explicit construction of solutions of type ($m$, $l=0$) is presented in section~\ref{largeK} for large values of the coupling constant $K$, where these solutions appear as stable phase waves travelling round the ring. Section~\ref{lowK} explores the low $K$ regime where the synchronized solutions with various values of $m$ and $l$ appear in succession. Detailed considerations for the case $N=3$ are presented in section~\ref{N=3} for a particular choice of the frequencies $\omega_i~(i=1,2,3)$ of the oscillators. Section~\ref{chaos} describes the phenomenon of intermittent chaos characterizing the system just before the appearance of the first synchronized solution as $K$ is made to increase from a low value. Finally, section~\ref{entrain} addresses the problem of entrainment of the synchronized solutions under a periodic external driving including that of stability of the entrained solutions. The entrainment scenarios for small and large values of the driving amplitude $A$ are briefly compared. The phenomenon of {\it entrainment of intermittent chaos} is also indicated on the basis of numerical results.

\section{Ring of oscillators with nearest neighbour coupling}\label{formulate}

\vn We start with $N$ number of oscillators with time-dependent phases $\theta_i~(i=1,2,\ldots,N)$, described by equations
\bs
\begin{align}
	\dot{\theta}_i=\omega_i+\frac{K}{2}\big( sin(\theta_{i+1}-\theta_i)+sin(\theta_{i-1}-\theta_i)\big),~(i=1,2,\ldots, N),\label{eq1a}
\end{align}

\nn where we have assumed nearest neighbour interactions and where the ring boundary condition is expressed in the form
\begin{align}
	\theta_{N+1}\equiv\theta_1,~\theta_0\equiv\theta_N.\label{eq1b}
\end{align}
\es

\nn In these equations, $\omega_i$ are the natural frequencies of the oscillators making up the ring, and $K$ is a coupling constant ($K>0$) which will be treated as a parameter characterizing the model. Following~\cite{kuramoto6}, we derive in the next section the algorithm for obtaining the synchronized solutions of the system and then work out the stable solutions explicitly in the large $K$ limit. 

\vn We look for synchronized solutions in the model, of the form
\begin{align}
	\theta_i(t)=\omega t+\phi_i,\label{eq2}
\end{align}
 
\nn where $\phi_i$ are time-independent phases, and $\omega$ stands for the common frequency of the oscillators in the synchronized state. Substituting in eq.~\eqref{eq1a}, one obtains
\begin{align}
	\omega=\omega_i+\frac{K}{2}\big( sin(\phi_{i+1}-\phi_i)+sin(\phi_{i-1}-\phi_i)\big),~(i=1,2,\ldots, N).\label{eq2A}
\end{align}

\nn Adding up all the $N$ equations in eq.~\eqref{eq2A}, one finds that $\omega$, the frequency of synchronized oscillation, has to be the {\it mean} of the oscillator frequencies:
\begin{align}
	\omega=\frac{1}{N}\sum_{i=1}^N\omega_i,\label{eq3}
\end{align}
 
\nn and the time-independent phases $\phi_i ~(i=1,2,\ldots,N)$ satisfy
\bs
\begin{align}
	\big( sin(\phi_{i+1}-\phi_i)+sin(\phi_{i-1}-\phi_i)\big)=-\frac{2}{K}\Delta_i~(i=1,2,\ldots,N).\label{eq4a}
\end{align}

\nn In this equation, $\Delta_i$ denotes the frequency deviation of the $i$th oscillator from the synchronized mean frequency,
\begin{align}
	\Delta_i=\omega_i-\omega~(i=1,2,\ldots,N),\label{eq4b}
\end{align}

\nn and the time-independent phases $\phi_i$ satisfy the ring boundary conditions
\begin{align}
	\phi_{N+1}\equiv\phi_1,~\phi_0\equiv\phi_N.\label{eq4c}
\end{align}
\es

\section{Synchronized solutions in the ring model}\label{synch}

\vn The basic equation~\eqref{eq4a} determining a synchronized solution involves the time-independent phases $\phi_i$ only through
\bs
\begin{align}
	p_i\equiv sin~\psi_i,~\psi_i\equiv \phi_i-\phi_{i-1},~(i=1,2,\ldots,N).\label{eq5a}
\end{align}

\nn in the form
\begin{align}
p_{i+1}-p_i=-\frac{2}{K}\Delta_i,~(i=1,2,\ldots,N),\label{eq5b}	
\end{align}

\nn where
\begin{align}
	p_{N+1}\equiv p_1.\label{eq5c}
\end{align}
\es

\nn Evidently, the $N$ equations~\eqref{eq5b} are not independent of one another since they add up to zero. Choosing $p\equiv p_1$ (say) as a parameter for the time being, all the other $p_i$'s are determined as
\bs
\begin{align}
	p_i=p+\frac{2}{K}s_i,~(i=1,2,\ldots, N),\label{eq6a}
\end{align}

\nn where
\begin{align}
s_i\equiv - \sum_{j=1}^{i-1}\Delta_j,~(i=2,\ldots,N),~s_1\equiv 0,\label{eq6b}	
\end{align}
\es

\nn and where, in equations~\eqref{eq6a}, the first equation results from the definitions of $p$ and $s_1$.

\vn With all the $p_i$'s so determined in terms of the parameters of the model (the coupling constant$K$ and the frequencies $\omega_i$) and of $p$, one can finally determine $p$ by noting that the phase differences $\psi_i$ have to satisfy
\begin{align}
	\sum_{i=1}^N \psi_i=2\pi m,\label{eq7}
\end{align}

\nn where $m$ is an integer, to be referred to as the winding number, on which more later. This equation involves $p_i$ through the relations $p_i=sin~\psi_i$ (see equation~\eqref{eq5a}), and hence determine $p$ by virtue of eq.~\eqref{eq6a}.  

\vn We note that $\psi_i$ is the inverse sine of $p_i$ which may lie in either of the two intervals $[-\frac{\pi}{2},+\frac{\pi}{2})$ and $[\frac{\pi}{2},\frac{3\pi}{2})$, i.e., with reference to the unit circle on which angles are depicted, either on the right half-circle or the left half-circle. For any given $x$ in the range $-1\leq x\leq +1$, we denote these two values of the inverse sine of $x$  as $\alpha(x)$ and $\beta(x)$ respectively. Here the end-point $\frac{3\pi}{2}$ is not identified with $-\frac{\pi}{2}$ since it may arise as a limiting value of an angle belonging to the left half circle. Thus, the equation~\eqref{eq7} determining $p$ (and hence a synchronized solution for the system) is finally seen to be of the form
\bs
\begin{align}
	\big(\alpha(p+\frac{2}{K}s_{i_1})+\cdots+\alpha(p+\frac{2}{K}s_{i_{N-l}})\big)+\big( \beta(p+\frac{2}{K}s_{j_1})+\cdots +\beta(p+\frac{2}{K}s_{j_l})\big)=2m\pi,\label{eq8a}
\end{align}

\nn where $l$ is any integer ranging from $0$ to $N$ and where it is understood that, for $l=0$, the second group of terms in the left hand side of the above equation is absent while, for $l=N$, the first group of terms is absent. Here $i_1,\ldots, i_{N-l}$ and $j_1,\ldots, j_l$ are two disjoint subsets of the set of integer suffixes $1,\ldots, N$. A synchronized solution is thus characterized by the integers $m$ and $l$, where $l$ denotes the number of phase differences $(\psi_i)$ that belong to the interval $[\frac{\pi}{2},3\frac{\pi}{2})$ (i.e., to the left half of the indicator circle).

\vn For the particular case of a ($m$,$0$) type solution, the above equation determining $p$ assumes the form
\begin{align}
	\sum_{i=1}^N \alpha(p+\frac{2}{K}s_i)=2m\pi.\label{eq8b}
\end{align}
\es

\nn We choose the number of oscillators $N$ and the frequencies $\omega_i$ (and hence $s_i,~i=1,\ldots,N$) as given parameters for a ring of oscillators, and treat the coupling constant $K$ as a variable parameter, and we address the task of finding the synchronized solutions of the system for various possible values of $K$. Any such synchronized solution corresponds to specific values of the integers $l(0\leq l\leq N)$ and $m$ and to a certain subset ${j_1,\ldots j_l}$ of the set ${1,\ldots, N}$ (the complementary subset ${i_1,\ldots,i_{N-l}}$ is automatically determined). For given $m$ and $l~(1\leq l\leq N)$, there may be several different solutions of type ($m,~l$) depending on which of the $N$ phase differences $\psi_i~{i=1,\ldots,N}$ are in the left half-circle and which in the right half-circle. 

\vn For a given $l$, since the upper and lower limits of the phase differences are known, one can determine the range over which the integer $m$ can vary, where one finds
\bs
\begin{align}
	-\frac{N}{4}+\frac{l}{2}< m<\frac{N}{4}+\frac{l}{2},\label{eq9a}
\end{align}

\nn where the equality signs are not included since these can arise only for a degenerate solution with all the $\psi_i$'s equal in the special case when all the oscillators have the same frequency.
 
\vn In particular, in the case of a ($m,0$) type solution, $m$ can be any interger in the range
\begin{align}
	-\frac{N}{4}< m < \frac{N}{4}.\label{eq9b}
\end{align}
\es

\nn The integer $m$ is referred to as the winding number characterizing the solution since, starting from the phase for any particular oscillator, say, $\phi_N$ and following the rotations on the unit circle by the successive phase differences $\psi_1,\psi_2,\ldots$, one comes back to the same oscillator after $m$ windings of the circle.

\section{Stability of a synchronized solution}\label{stability}

\nn For any given synchronized solution determined as above, a rigid rotation of all the phases $\phi_i~(i=1,\ldots,N)$ on the circle by an arbitrary angle $\Phi$, corresponding to the transformation $\phi_i\rightarrow\phi_i+\Phi$, does not alter the phase differences $\psi_i$. In other words, a synchronized solution is actually a member of a one-parameter family of solutions defined by rigid rotations. This immediately goes to show that the synchronized solutions are all neutrally stable, i.e., the matrix ($M$) describing the linearized equation obtained from eq.~\eqref{eq1a} always possesses at least one zero 
eigenvalue.

\vn Defining $q_i\equiv cos\psi_i$, for the synchronized solution under consideration, the elements $M_{ij}~(1\leq i,j \leq N)$ of this matrix are seen to be given by
\begin{align}
	M_{ii}=-\frac{K}{2}(q_i+q_{i+1}),~M_{i,i+1}=\frac{K}{2}q_{i+1},~ M_{ij}=M_{ji},~(1\leq i,j \leq N),\label{eq10}
\end{align}

\nn where, once again, the indices are to be interpreted in accordance with the ring boundary conditions ($N+1\equiv 1$, $0\equiv N$), and where all elements not specified by~\eqref{eq10} are zero. For instance, with $N=5$, the stability matrix is  
\begin{align}
	M=\frac{K}{2}\begin{pmatrix}-(q_1+q_2) & q_2 & 0 & 0 & q_1\\q_2 & -(q_2+q_3) & q_3 & 0 & 0\\0 & q_3 & -(q_3+q_4) & q_4 & 0\\ 0 & 0 & q_4 & -(q_4+q_5) & q_5\\q_1 & 0 & 0 & q_5 & -(q_5+q_1)\end{pmatrix}.\label{eq11}
\end{align}

\nn It is easy to see that, for any $N$, this matrix possesses a zero eigenvalue regradless of the values of $q_1,\ldots, q_N$ since, for the column $x=(1,1,\ldots, 1)^{\rm T}$, one has $Mx=0$.

\vn Moreover, in accordance with Greshgorin's theorem~\cite{rogge1,golub}, the following results hold: (a) if all the $\psi_i$'s belong to the right half circle, then all eigenvalues of $M$ are less than or equal to zero, in which case the synchronized solution under consideration is orbitally stable or orbitally neutrally stable, i.e., in other words the ($m$, $0$) type solutions are, in general, orbitally stable; (b) if, on the other hand, all the $\psi_i$'s are in the left half circle, the solution under consideration is orbitally unstable or, at most, orbitally neutrally stable, i.e., ($m$, $l=N$) type solutions are, in general, orbitally unstable.

\vn In addition, based on Greshgorin's theorem, the following result can be stated by employing a continuity argument: a ($m$, $1$) type solution, merging with a ($m$,0) type solution at some particular value of $K$ (see below), can remain orbitally stable as $K$ is made to vary continuously up to a certain limiting value.  
\section{Phase waves for large $K$}\label{largeK}

\vn For sufficiently large $K$, for which all the terms $\frac{2}{K}s_i ~(i=1,\ldots,N)$ are small and $p$ is bounded away from $\pm 1$ for a solution of type ($m$, $0$) (see below), one can make the expansion
\begin{align}
	sin^{-1}(p+\frac{2}{K}s_i)\approx sin^{-1}p+\frac{2}{K\sqrt{1-p^2}}s_i,\label{eq12}
\end{align}

\nn where higher order terms in the expansion are ignored.

\vn Thus, from eq.~\eqref{eq8b} the value of $p$ for a ($m$,$0$) type solution at large $K$ works out to 
\begin{align}
	p\approx sin \frac{2m\pi}{N}-\frac{2}{NK}\sum_{i=1}^N s_i,\label{eq13}
\end{align}

\nn where, once again, only terms of the order of $\frac{1}{K}$ are retained. This gives the result
\bs
\begin{align}
	\psi_i=sin^{-1}(p+\frac{2}{K}s_i)=\frac{2m\pi}{N}-\frac{2}{K}(s-s_i)sec(\frac{2m\pi}{N}),\label{eq14a}
\end{align}

\nn where
\begin{align}
	s\equiv \frac{1}{N}\sum_{i=1}^N s_i.\label{eq14b}
\end{align}
\es

\nn In the limit $K\rightarrow\infty$, one therefore obtains the synchronized ($m$,$0$) type solution in the form of a one-parameter family
\begin{align}
	\theta_i=\omega t+ \frac{2m\pi}{N}i+\delta,\label{eq15}
\end{align}

\nn where the parameter $\delta$ stands for the constant part of the phase (independent of $t$ and $i$), which can be chosen arbitrarily. Evidently, such a solution is in the nature of a phase wave running round the ring of oscillators, corresponding to a uniform phase advance between successive oscillators. Synchronized solutions for lower values of $K$, however, do not appear as phase waves since these are characterized by a non-uniform phase advance.

\vn Incidentally, the range of variation of $m$ given by~\eqref{eq9b} shows that $p$ (eq.~\eqref{eq13}) is indeed bounded away from $\pm 1$ as we assumed to start with. 

\section{Synchronised solutions: the low-$K$ scenario}\label{lowK}

\nn Continuing with the ($m$,$0$) type solutions for lower values of $K$, we will now see that a necessary condition for the existence of such a solution is that the coupling constant $K$ is to be larger than a certain minimum value $K_0$. For $K>K_0$, a ($m$,$0$) type solution appears at a certain threshold value $K_0^{(m)}$, which continues as a stable solution at larger values of $K$ till, at a sufficiently large value of $K$ the solution assumes the form of a phase wave given by~\eqref{eq14a}.

\vn Since the phase differences are given by $sin\psi_i=p+\frac{2}{K}s_i$, a necessary condition for the existence of a synchronized solution is that $-1<p+\frac{2}{K}s_i<1$ for all $i=1,\ldots,N$. In other words, denoting by $x$ and $y$ the maximum and minimum values in the set $\{s_1,\ldots,s_N\}$, $p$ and $K$ must satisfy the conditions
\begin{align}
	p+\frac{2}{K}x ~\leq 1,~p+\frac{2}{K}y \geq -1.\label{eq16}
\end{align}

\nn Thus, defining 
\begin{align}
	p_+(K)\equiv 1-\frac{2}{K}x,~p_-(K)=-1-\frac{2}{K}y,\label{eq17}
\end{align}

\begin{figure}[htb]
	\centering
		\includegraphics[width=0.8\textwidth]{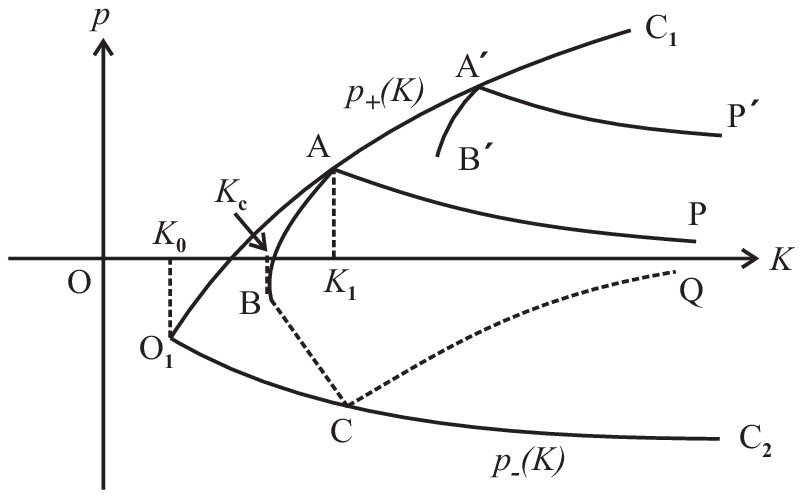}
		\caption{Graphs showing the variation of $p_-(K)$ and $p_+(K)$ with $K$ (schematic); the two graphs intersect at $p=p_0,~K=K_0$; the values of $p$ and $K$ for any synchronized solution has to correspond to a point lying in between these two curves; thus, for the existence of a synchronized solution, $K$ has to be larger than $K_0$.}
	\label{kura1}
\end{figure}

\nn the possible values of $p$ and $K$ for which synchronized solutions exist correspond to points lying in between the graphs for $p_+(K)$ against $K$, and $p_-(K)$ shown schematically in fig.~\ref{kura1}, where the point of intersection of the two graphs corresponds to
\begin{align}
K=K_0\equiv x-y,~p=p_0\equiv -\frac{x+y}{x-y}.\label{eq18}
\end{align}

\nn As an example of how a ($m$, $0$) type solution emerges as $K$ is made to increase from the value $K_0$, we consider the emergence of the ($m=0$,$l=0$) solution. This solution is determined by the relation
\begin{align}
	\sum_{i=1}^N \alpha (p+\frac{2}{K}s_i)=0,\label{eq19}
\end{align}
 
\vn where, recall that $\alpha(x)$ stands for that value of $sin^{-1}x$ which lies in the range $-\frac{\pi}{2}\leq x\leq\frac{\pi}{2}$ (the inclusion of the end-point $\frac{\pi}{2}$ does not involve any inconsistency since, as we see below, this value is relevant only at a point where a solution of a given type passes continuously into one of a different type). 

\vn Evaluating the left hand side of eq.~\eqref{eq19} with $p=p_0$ and $K=K_0$ , one obtains the expression 

\begin{align}
P\equiv\sum_{i=1}^N \alpha\big(\frac{2s_i-(x+y)}{x-y}\big).\label{eq19N} 
\end{align}

\vn If this expression is negative then, considering the two graphs of $p_-(K)$ and $p_+(K)$ referred to above, the equation~\eqref{eq19} can be satisfied only if ($p,~K$) lies on the graph for $p_+(K)$ while, conversely, if $P$ is positive then eq.~\eqref{eq19} can be satisfied only on the graph of $p_-(K)$. This is because $p_+(K)$ increases monotonically with increasing $K$ while $p_-(K)$ decreases monotonically.

\vn Let us assume, for the sake of illustration, that the frequencies of the oscillators are such that $P<0$. Then, starting from the point $(K_0,p_0)$, as we move along the upper graph, a point A($K^{(0)},p_+(K^{(0)})$) will be encountered  at which eq.~\eqref{eq19} will be satisfied. The value $K=K^{(0)}$ (with a slight change in notation, $K^{(0)}$ is shown as $K_1$ in fig.~\ref{kura1} for the sake of convenience) then corresponds to the emergence of the ($0,0$) type solution since, for $K<K^{(0)}$, at least one of the expressions $p+\frac{2}{K}s_i~(i=1,\ldots,N)$ will be greater than $+1$. On the other hand, for a $K$ satisfying $K>K^{(0)}$, one will have a $(0,0)$ type solution with $p(K)$ satisfying~\eqref{eq19}. Here we have used the abbreviated notation $K^{(0)}$ for $K_0^{(0)}$, where $K=K_0^{(m)}$ corresponds to the appearance of a $(m,0)$ type solution.

\vn What is interesting to note in this connection is that, when $K$ is slightly {\it less} than $K^{(0)}$ one can, by continuity, have a solution to eq.~\eqref{eq8a} with $l=1$, where now the largest of the phase differences $\psi_i$ will be slightly larger than $\frac{\pi}{2}$. In other words, a $(m=0,l=1)$ solution transforms to a $(m=0,l=0)$ solution at $K=K^{(0)}$, the latter being the value of $K$ for which the largest of the phase differences for either of the two solutions assumes the value $\frac{\pi}{2}$. 

\vn Of the two synchronized solutions that meet at $K=K^{(0)}$, the ($0,0$) solution is orbitally stable by Greshgorin's theorem since all the $q_i$'s in the linear stability matrix $M$ (refer to sec.~\ref{stability}) are positive. For $K$ close to $K^{(0)}$, the associated ($0,1$) type solution is {\it also} stable by continuity, since the two sets of phase differences $\{\psi_i~(i=1,\ldots,N)\}$ are close to each other.

\vn As $K$ is made to increase away from $K^{(0)}$, the corresponding values of $p$ for the $(0,0)$ type synchronized solution trace out a graph depicted schematically in fig.~\ref{kura1} (recall that this illustration assumes $P<0$, where $P$ is given by the expression~\eqref{eq19N}), which is drawn with a solid line to indicate that it is an orbitally stable solution. Starting from the point A ($K=K^{(0)},p=p_+(K^{(0)})$), the graph approaches asymptotically the value $p=0$ in accordance with~\eqref{eq13} for $K\rightarrow\infty$. 

\vn As indicated above, the corresponding graph for the $(0,1)$ type solution is seen to start at the same point A. Considering for the sake of concreteness, the case of $N=3$ (see sec.`\ref{N=3}), the graph depicting the variation of $p$ with $K$ for the $(0,1)$ solution is seen to include a point B where it shows a {\it backbending}. At the point B (for which $\frac{dK}{dp}=0$), there occurs a merger of a stable and an unstable branch of the $(0,1)$ type solution. Thus, in other words, the $(0,1)$ solution is stable for values of $K$ (and $p$) in between the points A and B, while it becomes unstable at B and the $p$-$K$ graph corresponding to the unstable $(m=0,l=1)$ solution continues up to C, the latter being located on the graph of $p_-(K)$ against $K$. As $K$ is made to increase beyond the point C, on the other hand, one obtains yet another solution, namely one of type $(m=1,l=2)$. The latter is an unstable solution that continues up to $K\rightarrow\infty$, the variation of $p$ with $K$ being shown schematically in the figure.

\vn {\it In summary}, then, fig.~\ref{kura1} depicts the graphs of $p_+(K)$ and $p_-(K)$ as functions of $K$, with reference to which the solutions of various types characterized by the integers $m,~l$ can be described. Each synchronized solution originates on either of the two curves (O$_1$C$_1$, O$_1$C$_2$ in the figure) and the variation of $p$ with $K$ for the solution corresponds to a curve such AP, ABC or CQ. While the curve AP corresponds to the $(m=0,l=0)$ solution, other solutions of type $(m,0)$, for the various other possible values of $m$ (refer to formula~\eqref{eq9b}) also correspond to similar curves. The figure shows schematically another such curve A$'$P$'$ where A$'$ is the point at which a stable $(m,1)$ type solution (for which the $p$-$K$ curve extends along A$'$B$'$) transforms into one of type $(m,0)$. As already mentioned, all these $(m,0)$ type solutions are of the nature of stable phase waves for large values of $K$, the phases of the oscillators on the ring for such a solution being given by eq.~\eqref{eq15}. Meanwhile, solutions for various other values of $l$ are also described in terms of their own $p$-$K$ curves. 

\vn The following section includes a detailed consideration of the solutions for a particular case with three oscillators on a ring where we describe all the possible synchronized solutions. In the process we identify the critical value $(K_{\rm c})$ of $K$ (corresponding to the point (C) of backbending of the curve ACB in fig.~\ref{kura1}) for the onset of synchronization. For $K$ slightly below $K_{\rm c}$, the system exhibits {\it intermittent chaos}. 

\section{A ring with three oscillators}\label{N=3}

\vn We consider a ring of $N=3$ oscillators with frequencies $\omega_1,\omega_2,\omega_3$, for which we assume, for the sake of concreteness, $\omega_1=1$, $\omega_2=-3$, and $\omega_3=2$, in arbitrary units, corresponding to which one has $\omega=0$, $s_1=0$ (by definition), $s_2=-1(=y),~s_3=2(=x)$, .

\vn Synchronized solutions for this ring can be denoted symbolically by indicating whether the successive phase differences $\psi_i (i=1,2,3)$ belong to the interval $[-\frac{\pi}{2},\frac{\pi}{2}]$ (which we indicate by the symbol 'r', corresponding to the right half circle) or to $[\frac{\pi}{2},\frac{3\pi}{2}]$ (symbol 'l', corresponding to the left half circle) and, in adition, by the value of $m$ (the value of $l$ is determined by the number of 'l' symbols in the string characterizing the intervals to which the successive phase difference belong). For instance, a 'rrr(0)' solution corresponds to all the three phase differences being of type 'r' (hence implying $l=0$), with the sum of the three being $0$. Similarly, a 'rll(1)' solution corresponds to $\psi_1$ being of type 'r' and $\psi_2$, $\psi_3$ being of type 'l' (hence $l=2$), with the sum of the three being $2\pi$. The two intervals mentioned above include common end-points since these correspond to the points of transition of solutions of one type into another, such as the points A and C in fig.~\ref{kura1}.

\vn Making use of this notation one can make up a list of the possible synchronized solutions for the system under consideration. The possible values of $l$ being $l=0,1,2,3$, the corresponding values of $m$ in accordance with the formula~\eqref{eq9a} are as follos

\begin{align}
	l&=0: m=0,\nonumber\\
	l&=1: m=0,1,\nonumber\\
	l&=2: m=1,\nonumber\\
	l&=3: m=1,2.\label{eq20}
\end{align}
 
\nn The above list then includes a total of twelve possible solutions which we label as rrr(0), rrl(0), rrl(1), rlr(0), rlr(1), lrr(0), lrr(1), rll(1), lrl(1), llr(1), lll(1), and lll(2).

\vn Considering, to begin with, the rrr(0) synchronized solution for the sake of concreteness, and following notations as in sec.~\ref{synch} one finds that, for any given $K$, there exists a $p (=sin~\psi_1)$ corresponding to such a solution, provided the following equation is satisfied,

\begin{align}
	sin^{-1}(p)+sin^{-1}(p+uy)+sin^{-1}(p+ux)=0,\label{eq21}
\end{align}
 
\nn Here we use the notation $u=\frac{2}{K}$ for the sake of brevity and note that the function $sin^{-1}$ denotes the principal value of the inverse sine, which corresponds to an angle lying in the range $[-\frac{\pi}{2},\frac{\pi}{2}]$ (this was denoted by the symbol $\alpha$ in eq.~\eqref{eq8a}, while the symbol $\beta$ was used to denote the inverse sine lying in the range $[\frac{\pi}{2},\frac{3\pi}{2}]$; the two function branches agree at the argument value $+1$ ). Recall that, once $p$ is determined from the above equation for any given value of $K$ (i.e., equivalently, of $u$), the phase angle differences $\psi_i=\phi_i-\phi_{i-1} (i=1,2,3)$ are obtained as, respectively, the three terms on its left hand side, corresponding to which one obtains a one-parameter family of solutions for the phase angles $\phi_i$ themselves (see sec.~\ref{stability}), and a one-parameter family of synchronized solutions is thereby obtained from equations~\eqref{eq2},~\eqref{eq3}. In the following, while speaking of a synchronized solution (such as the rrr(0) solution under consideration) we will actually mean the corresponding one-parameter family.

\vn As mentioned in sec~\ref{lowK}, synchronized solutions do not exist for $K$ less than a certain threshold value $K_0$, corresponding to the point of intersection of the graphs of $p_+(K)$ and $p_-(K)$(refer to eq.~\eqref{eq17} which, with our choice of parameters, is given by $K_0=3, p_+=p_-=p_0(\rm{say})=-\frac{1}{3}$. As $K$ is made to increase beyond $K_0$, the rrr(0) solution first makes its appearance at 

\begin{align}
K=K_1({\rm{say}})=\frac{2(2x^2+y^2-2xy)}{2x-y+\sqrt{2x^2-2xy}}, p=p_1({\rm{say}})=\frac{y^2-xy-x\sqrt{2x^2-2xy}}{2x^2+y^2-2xy},\label{eq21A} 
\end{align}

\nn where the point $(K_1,p_1)$ lies on the graph of $p_+(K)$ (note the slight change in notation compared to sec.~\ref{lowK} where the symbol $K^{(0)}$ was used in place of $K_1$). This means that, for $K=K_1$, the equation~\eqref{eq21} is satisfied with $p=p_1$, for which $\psi_3=\frac{\pi}{2}$, the latter being one end point of the range $[-\frac{\pi}{2},\frac{\pi}{2}]$. Since this is also an end point of the range $[\frac{\pi}{2},\frac{3\pi}{2}]$, it is evident that $(K_1,p_1)$ also satisfies the equation

\begin{align}
	sin^{-1}(p)+sin^{-1}(p+uy)+(\pi-sin^{-1}(p+ux))=0,\label{eq22}
\end{align}

\begin{figure}[htb]
	\centering
		\includegraphics[width=0.8\textwidth]{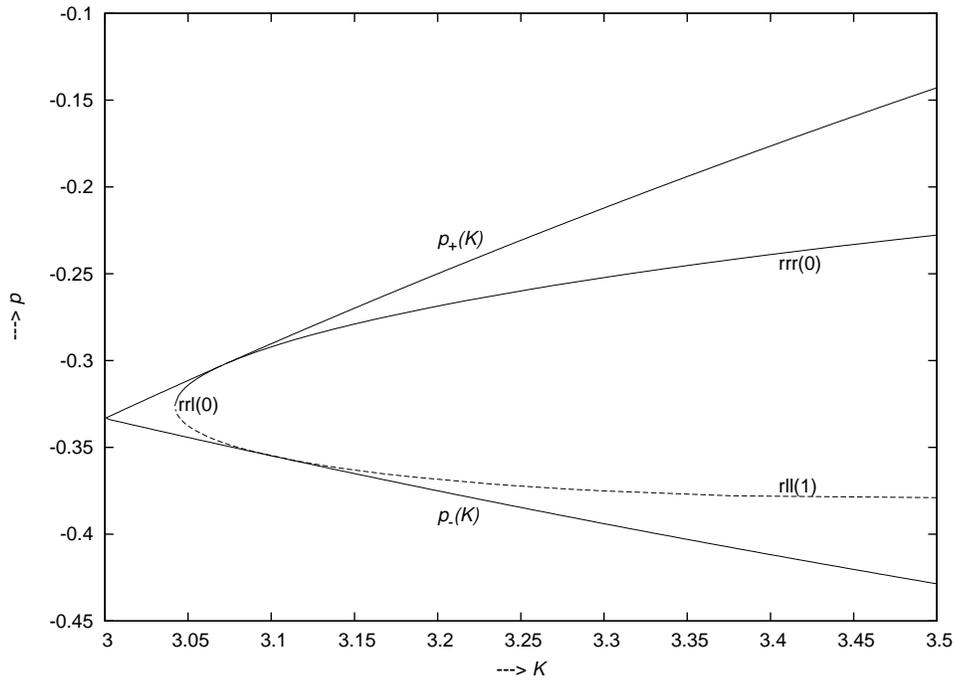}
		\caption{$p$-$K$ curves for the rrr(0), rrl(0) and rll(1) synchronized solutions; the point of backbending for the rrl(0) curve corresponds to the point C in fig.~\ref{kura1}, where a stable and an unstable branch originate by a tangent bifurcation.}
	\label{kura2}
\end{figure}

\nn which is precisely the equation determining a rrl(0) solution. However, the $p$-$K$ curve for the rrl(0) solution extends to the left of A (fig.~\ref{kura1}), i.e., in other words, as $K$ is made to increase from $K_0$, the rrl(0) solution gets transformed into rrr(0) as $K$ crosses $K_1$. This $p$-$K$ curve, describing the history of the rrl(0) solution has a point of backbending (the point B in fig.~\ref{kura1}, see also fig.~\ref{kura2} obtained numerically) where there occurs a change in stability of this solution so that the segment CB corresponds to an unstable rrl(0) solution. 

\vn The eigenvalues of the stability matrix $M$ are seen to be
\begin{align}
	\lambda_1=0,~\lambda_{2,3}=\frac{K}{2}\big(-(q_1+q_2+q_3)\pm\sqrt{(q_1^2+q_2^2+q_3^2-q_1q_2-q_2q_3-q_3q_1)}~\big),\label{eq23}
\end{align}

\nn and thus the value $(K_{\rm c})$ of $K$ for the point B corresponds to the condition
\begin{align}
	q_1q_2+q_2q_3+q_3q_1=0.\label{eq24}
\end{align}

\nn For our choice of the parameters $\omega_1,~\omega_2,~\omega_3$, the condition~\eqref{eq24} corresponds to
\begin{align}
	K_{\rm c}=3.04224~(\rm{approx}).\label{eq25}
\end{align}

\nn Looking at fig.~\ref{kura1}, one arrives at the important cnclusion that this is the critical value of $K$ below which there exists no synchronized solution for the system under consideration while, as $K$ is made to increase past $K_{\rm c}$, there appears a pair of rrl(0) solutions by tangent bifurcation, of which one is a stable synchronized solution while the other is an unstable one. For $K$ slightly below this critical value ($K_{\rm c}$) the system exhibits {\it intermittent chaos} (see section~\ref{chaos}).

\vn As $K$ is made to increase beyond $K_{\rm c}$, the stable branch of the $p$-$K$ curve describing the rrl(0) solution terminates on the $p_+(K)$-$K$ curve (the point A on the branch O$_1$C$_1$ in fig.~\ref{kura1}), and the rrl(0) solution transforms to a rrr(0) one described by its own $p$-$K$ curve (branch AP in 
fig.~\ref{kura1}. As already mentioned, in the large-$K$ regime this $p$-$K$ curve tends to the limit $\psi_1=\psi_2=\psi_3=0$ (refer to eq.~\eqref{eq14a}) which thus corresponds to a phase-locked solution for the system under consideration.

\vn Meanwhile, for $K$ beyond $K_{\rm c}$ the unstable branch of the rrl(0) solution terminates on the $p_-(K)$ curve (point C in fig.~\ref{kura1}) at
\begin{align}
K=K_2({\rm{say}})=\frac{2(x^2+2y^2-2xy)}{x-2y+\sqrt{2y^2-2xy}}, p(K_2)=p_2({\rm{say}})=\frac{-x^2+xy-y\sqrt{2y^2-2xy}}{x^2+2y^2-2xy},\label{eq21A} 
\end{align}

\nn where the rrl(0) solution gets transformed into a rll(1) one, an unstable synchronized solution for the system. The latter continues, for larger values of $K$ up to $K\rightarrow\infty$, when it tends to an unstable synchronized solution with $\psi_1=0,~\psi_2=\pi,~\psi_3=\pi$.

\begin{figure}[htb]
	\centering
		\includegraphics[width=0.8\textwidth]{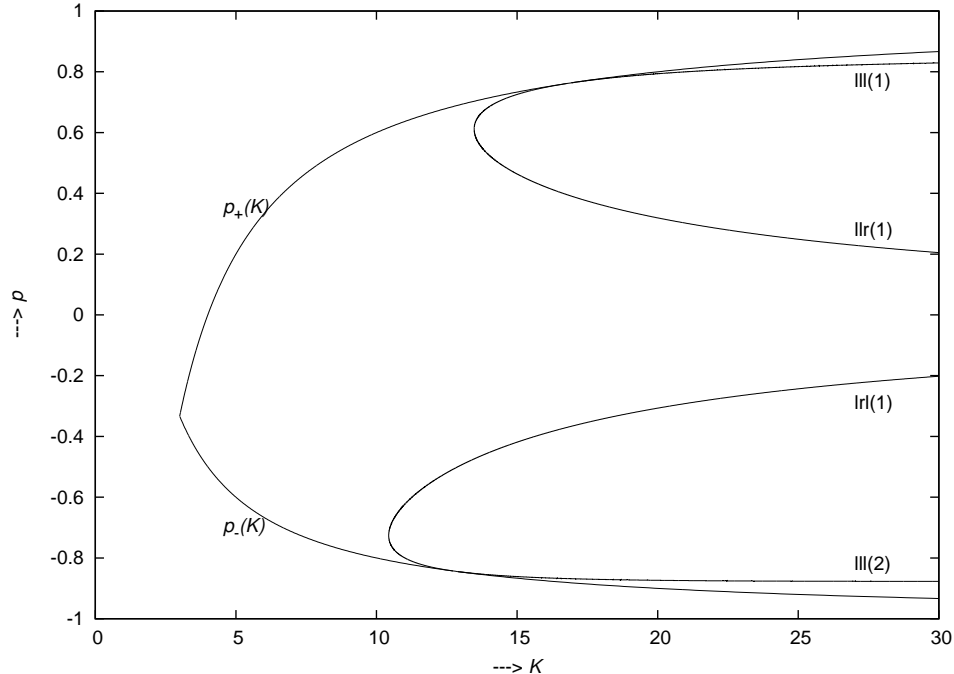}
		\caption{$p$-$K$ curves for the llr(1), lll(1), lrl(1), and lll(2) synchronized solutions; the point of backbending on either of the llr(1) and lrl(1) solutions is analogous to that on the rrl(0) curve of fig.~\ref{kura2}, where one eigenvalue of the stability matrix changes sign.}
	\label{kura3}
\end{figure}

\vn Figures~\ref{kura2},~\ref{kura3} show the numerically computed $p$-$K$ curves for the rrr(0), rrl(0), rll(1)type solutions, and the llr(1), lll(1), lrl(1) and lll(2) type solutions respectively. for the system with parameter values mentioned above. One observes that the lll(2) solution gets transformed into a lrl(1) solution and, similarly, a ll1(1) solution gets transformed into a llr(1) solution in a manner analogous to the way a rrr(0) solution changes over to a rrl(0) one. Among these, the lll(1) and lll(2) solutions are in the nature of synchronized phase waves on the ring for large $K$, with $\psi_1=\psi_2=\psi_3=\frac{2\pi}{3}$ and $\psi_1=\psi_2=\psi_3=\frac{4\pi}{3}$ respectively. Moreover, there occur points of backbending for the $p$-$K$ curves describing the llr(1) and lrl(1) solutions, analogous to a similar backbending for the rrl(0) solution as seen above. These points once again correspond to the condition $q_1q_2+q_2q_3+q_3q_1=0$ (refer to eq.~\eqref{eq24}), but now one of the two eigenvalues $\lambda_{2,3}$ remains positive while the other passes through zero, as a result of which the solution under consideration remains unstable on either side of the point of backbending.  

\vn While figures~\ref{kura2},~\ref{kura3} depict the history of seven of the twelve possible synchronized solutions mentioned above, the remaining five (lrr(0), lrr(1), rlr(0), rlr(1), and rrl(1) ) happen to be ruled out for the particular situation under consideration in this section, since in each case, the existence of a solution to the equation determining $p$ as a function of $K$ (a particular instance of the general equation`\eqref{eq8a}) may be seen to be inconsistent with the requirement $-1\leq p\leq +1$ for $K_0<K<\infty$. 

\vn In other words, for a ring with $N=3$ oscillators with the parameter values ($\omega_1,\omega_2,\omega_3$) chosen as above, there exist only seven synchronized solutions, of which one exists within a finite range of the parameter $K$ while the other six persist up to large values of $K$. Among the latter six, only one is a stable synchronized solution with $\phi_1=\phi_2=\phi_3$ (phase-locking) while the others are all unstable.

\vn In a recent work on a closely related model~\cite{mehta1}, Mehta and Kastner have worked out fixeed points of a certain functional (the lattice Landau gauge fixing functional) in the context of lattice gauge theory that correspond to the synchronized solutions of the Kuramoto ring considered here, and their results also relate to the stability characteristics of the synchronized solutions. The extension to two dimensional systems has also been considered~\cite{mehta2}.

\section{Intermittent chaos}\label{chaos}

\begin{figure}[htb]
	\centering
		\includegraphics[width=0.8\textwidth]{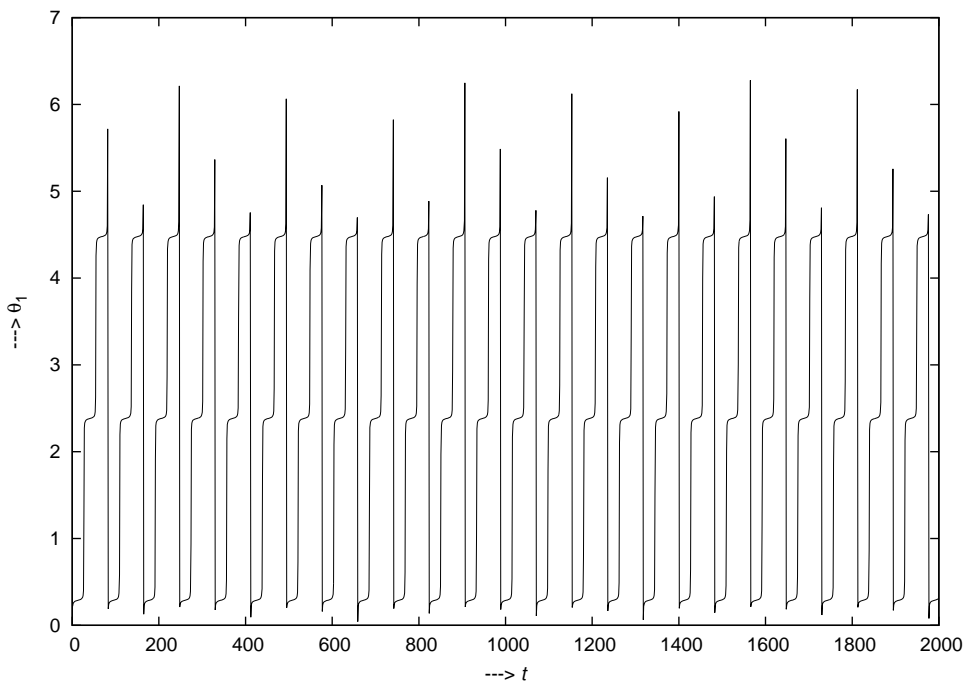}
		\caption{Intermittent chaos for $K$ slightly below the critical value $K_{\rm c}(\approx 3.04224)$; alteration of laminar and turbulent phases as seen in a numerical computation of the evolution of the phase angle $\theta_1$ with $K=K_{\rm c}-0.00390625$; the initial condition ($\theta_i(t=0)=\phi_i^{(0)}~(i=1,2,3)$) has been chosen as explained in the text; the laminar phase corresponds to values of $\theta_1$ close to $\phi_1^{(0)}$, $\phi_1^{(0)}+\frac{2\pi}{3}$, $\phi_1^{(0)}+\frac{4\pi}{3}$.}
	\label{kura4}
\end{figure}

\nn Fig.~\ref{kura4} depicts the alteration of laminar and turbulent phases in the intermittennt chaos (\cite{intermit}; the time evolution of the phase angles in the laminar phase may be approximated by a discrete mapping) for $K$ slightly below $K_{\rm c}$, where we look at a trajectory initiated close to the point on the 3-torus with $\psi_1,~\psi_2,~\psi_3$ corresponding to the rrl(0) solution at the point of the tangent bifurcation (at $K=K_{\rm c}$), and where we make the choice $\phi_1=0$ from among the one-parameter family of solutions corresponding to these values of $\psi_1,~\psi_2,~\psi_3$ (this gives $\phi_1=0,~\phi_2\approx -1.39002335,~\phi_3\approx 0.33238947$). 

\nn As seen in fig.~\ref{kura4}, the laminar phase alternates between values of $\theta_1$ close to $\phi_1^{(0)}$, $\phi_1^{(0)}+\frac{2\pi}{3}$, $\phi_1^{(0)}+\frac{4\pi}{3}$, $\phi_1^{(0)}$ being the initial value of $\theta_1$ chosen close to $0$ (the deviation from $0$ was chosen with a random number generator). This is explained by the fact that $\theta_1+\theta_2+\theta_3$ is a constant of motion for the system under consideration and , at the same time, a constant shift $\Phi$ in the three phase angles results in the same orbit. taking these two facts togethr, one finds that the constant shift $\Phi$ can be $\frac{2\pi}{3}$ or $\frac{4\pi}{3}$ for any given initial condition.

\vn It may be noted that, near the tangent bifurcation, the system behaves effectively as a one dimensional one since, among the three angles $\theta_1,~\theta_2,~\theta_3$, one is trivially eliminated by the condition $\frac{1}{3}(\theta_1+\theta_2+\theta_3=\omega t) $ (which, moreover, is zero for our choice of the parameters) following from equations~\eqref{eq1a},~\eqref{eq1b}, and then a centre manifold reduction leads to a single effective angular variable (say, $\Theta$), with the remaining, third variable being slaved to it (recall that, close to the tangent bifurcation, one of the three eigenvalues of the stability matrix is trivially zero while, among the remaining two, the relevant eigenvalue crosses through zero and the third eigenvalue remains negative). The equation of motion for the relevant angular variable $\Theta$ in the laminar phase can then be assumed to be of the standard form
\begin{align}
	\dot\Theta=\epsilon+\Theta^2,~(\epsilon=K_{\rm c}-K).\label{eq26}
\end{align}

\begin{figure}[htb]
	\begin{center}
		\includegraphics[width=0.8\textwidth]{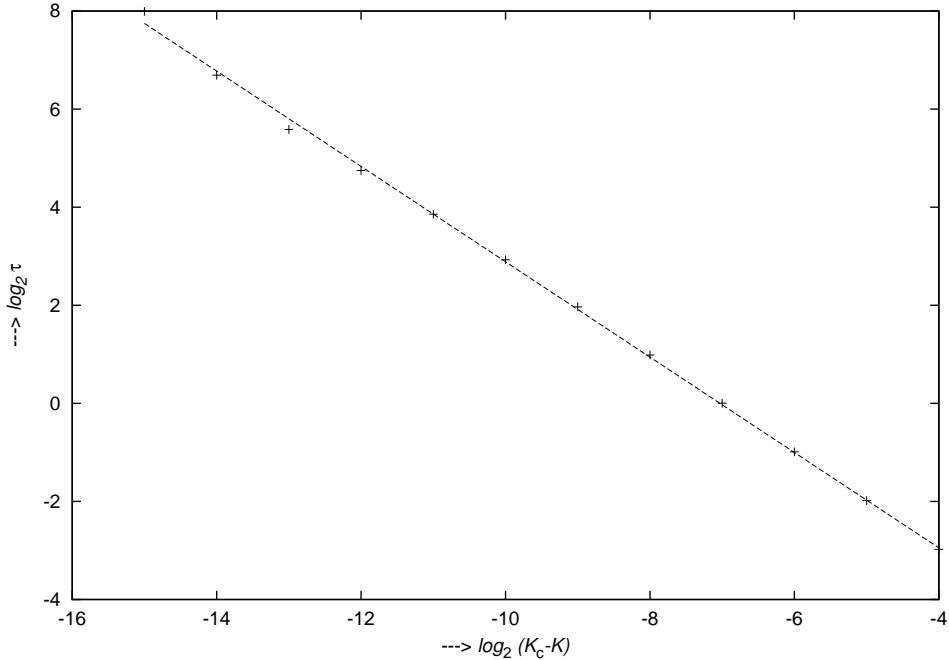}
		\end{center}
		\caption{Scaling of the duration ($\tau$) of the laminar phase; $\tau$ scales with $\epsilon=K_{\rm c}-K$ as $\epsilon^{\nu}$, where $\nu~(\approx -0.97)$ is seen to be close to the theoretical value $-1$.}
	\label{kura5}
\end{figure}

\nn As is evident from a dimensional argument, the duration of the laminar phase is then expected to scale with $\epsilon$ as $\epsilon^{-1}$. As seen in fig.~\ref{kura5}, the scaling exponent obtained numerically for the system under consideration is indeed close to $-1$.

\vn The above phenomenon of intermittent chaos is a particular instance of the general phenomenon of chaos in systems of coupled oscillators before the onset of synchronization (see, e.g.,~\cite{kuramoto5},~\cite{kuramoto8} cited above). In sec.~\ref{chao-entrain} below we include a few numerical results on our system where the intermittent chaos is {\it entrained} to a synchronized solution by means of an external periodic driving.
 
\section{Periodic driving: entrainment of synchronized oscillations}\label{entrain}

\subsection{Entrainment: a perturbative approach}\label{ent-perturb}

\nn We now include a periodic driving term of amplitude $A$ and frequency $\Omega$ in eq.~\eqref{eq1a} so as to obtain the system
\begin{align}
	\dot{\theta}_i=\omega_i+\frac{K}{2}\big( sin(\theta_{i+1}-\theta_i)+sin(\theta_{i-1}-\theta_i)\big)+A sin (\Omega t-\theta_i+\delta),~(i=1,2,\ldots, N; A>0)\label{eqA1}
\end{align}

\nn where $\delta$ denotes a constant determining the difference between the driving phase $(\Omega t+\delta)$ and the free-running phase angle (i.e., the phase angle in the absence of driving) characterizing the $i$th oscillator ($i=1,2,\ldots,N$) and where the ring boundary conditions are implied. Driven populations of oscillators with a global coupling between them have been considered in~\cite{ott5} which shows a rich pattern of behaviour for such systems, involving a number of bifurcation routes. In the present paper, we look for entrained solutions where the phases of the oscillators on the ring all vary periodically with the driving period $\Omega$ as
\begin{align}
	\theta_i=\Omega t +\phi_i +\gamma_i.\label{eqA2}
\end{align}

\nn Here $\phi_i$ denotes the constant part of the phase of the $i$th oscillator ($i=1,2,\ldots, N$) for a synchronized solution of the form~\eqref{eq2} which we take as a reference, calling the latter a {\it free-running} solution. The driving frequency $\Omega$ characterizing the entrained solution differs from the free-running frequency $\omega$ by $\Delta=\Omega-\omega$, which we refer to as the detuning. Finally, $\gamma_i$ ($i=1,2,\ldots,N$) are the deviations of the constant parts of the phases from the corresponding free-running values $\phi_i$.

\vn We assume that the driving amplitude and the detuning are sufficiently small, as a result of which there corresponds to the free-running solution an entrained solution that differs by a small extent from the former, and outline a perturbative procedure for obtaining the phase deviations $\gamma_i$ for the latter. Substituting the proposed solution~\eqref{eqA2} in~\eqref{eq1a} one obtains
\begin{align}
	sin(\phi_{i+1}-\phi_{i}+\gamma_{i+1}-\gamma_{i}) + sin (\phi_{i-1}-\phi_i+\gamma_{i-1}-\gamma_i)=\frac{2}{K}(\Omega-\omega_i-A sin (\delta-\phi_i-\gamma_i)).\label{eqA3}
\end{align}

\nn This system of equation is consistent only if
\begin{align}
	\sum_{i=1}^N(\Omega-\omega_i- A sin (\delta-\phi_i-\gamma_i))=0.\label{eqA4}
\end{align}

\nn Making use of this condition, an approximation scheme for working out the solution for sufficiently small $A$ and $\delta$ can be set up as follows.

\vn We first choose appropiate variables in place of $\gamma_i~(i=1,\ldots, N)$ by referring to the linearization of the left hand side of~\eqref{eqA3}. On making use of eq.~\eqref{eq2A}, this can be expressed in the form $M_1\gamma$, where $M_1$ is precisely $\frac{2}{K}$ times the stability matrix $M$ given by~\eqref{eq10} (with $\phi_i~(i=1,\ldots,N)$ corresponding to the synchronized solution under consideration), and where $\gamma$ stands for the column $\gamma=\{\gamma_1,\gamma_2,\ldots,\gamma_N\}^{\rm T}$. Since $M$ possesses a zero eigenvalue, a solution of eq.~\eqref{eqA3} by straightforward linearization and matrix inversion is not possible. If $T$ be the orthogonal matrix diagonalizing $M$, then the normal co-ordinates for the linearized problem (say, $\rho_1,\rho_2,\ldots,\rho_N$) are given by 
\begin{align}
	\rho=\{\rho_1,\ldots,\rho_N\}^{\rm T}=T^{\rm T}\gamma.\label{eqA5}
\end{align}

\nn If the eigenvalues of $M$ are denoted by $(\lambda_1=0,\lambda_2,\ldots,\lambda_N)$ (i.e., those of $M_1$ are $\frac{2}{K}\lambda_i~(i=1,2,\ldots,N)$), then the first column of $T$, corresponding to the eigenvector for the eigenvalue zero, is made up of $\frac{1}{\sqrt{N}}$'s as all the entries, and one has

\begin{align}
	\rho_1=\frac{1}{\sqrt{N}}\sum_{i=1}^N\gamma_i,\label{eqA6}
\end{align}

\vn where $\frac{1}{\sqrt{N}}$ is a normalization factor.

\vn The linearization of the left hand side of eq.~\eqref{eqA3} would then give (refer to equations~\eqref{eq5a},~\eqref{eq5b})
\begin{align}
	\Lambda\rho=T^{\rm T}\zeta,\label{eqA7}
\end{align}

\nn where $\zeta\equiv\{\zeta_1,\zeta_2,\ldots,\zeta_N\}^{\rm T}$, with
\begin{align}
	\zeta_i=\frac{2}{K}(\Delta-A sin (\delta-\phi_i-\gamma_i)),\label{eqA8}
\end{align}
 
\nn and $\Lambda$ is the diagonal form of $M_1$, with $\frac{2}{K}\lambda_i$ as the $i$th diagonal element.   
 
\vn Noting that the first row of $T^{\rm T}$ is just $\frac{1}{\sqrt{N}}\{1,1,\ldots,1\}^{\rm T}$, one finds that the first of the $N$ equations in~\eqref{eqA7} is an identity (in view of eq.~\eqref{eqA4}), which is why one has to retain the nonlinear terms in $\zeta$. Indeed, the choice of the new variables $\rho_i(i=1,2,\ldots, N)$ is a useful one because, as we see below, it allows us to make a clear separation between large and small components in the column $\rho$. It turns out that $\rho_1$ is large compared to the remaining components and includes a uniform phase translation for all the oscillators in the ring. In contrast to the other components of $\rho$, $\rho_1$ is appropriately solved for from the consistency relation eq.~\eqref{eqA4}. The latter can be written in the form (recall that $\sum_i \omega_i=N\omega$)

\begin{align}
\sum_{i=1}^{N}\big(\Delta-A sin (\Delta-\phi_i-(T\rho)_i)\big)=0~(\Delta=\Omega-\omega).\label{eqA9}
\end{align}

\vn The basic idea underlying our approximation scheme is that, for sufficiently small $A$ and $\Delta$, $\rho_2,\ldots,\rho_N$ remain small, being obtained from the $N-1$ equations of~\eqref{eqA7} with the first equation omitted, while $\rho_1$, which is expected to be large compared to $\rho_2,\ldots,\rho_N$, is to be solved for from eq.~\eqref{eqA9}. Accordingly, we first define
\begin{align}
	\gamma^{(0)}=\big(T\rho\big)_{\rho_2=\ldots=\rho_N=0},\label{eqA10}
\end{align}

\nn i.e., expressing $\gamma$ in terms of the $\rho_i$'s, we retain only $\rho_1$, ignoring the rest of the normal co-ordinates. Noting, however, that the first column of $T$ is just $\frac{1}{\sqrt{N}}\{1,\ldots,1\}^{\rm T}$, we obtain (for $i=1,2,\ldots,N$)
\begin{align}
	\gamma^{(0)}_i=\frac{\rho_1}{\sqrt{N}}.\label{eqA11}
\end{align}
 
\nn The required entrained solution is then obtained in the following steps.

\vn Substituting $\gamma^{(0)}=\frac{1}{\sqrt{N}}\{\rho_1,\ldots,\rho_1\}^{\rm T}$ for $T\rho$ in eq.~\eqref{eqA9}, we solve for $\rho_1$ from eq.~\eqref{eqA9}, i.e.from
\begin{align}
\sum_{i=1}^{N}\big(\Delta-A sin (\delta-\phi_i-\frac{\rho_1}{\sqrt{N}})\big)=0.\label{eqA12}
\end{align}

\nn Evidently, the solution for $\rho_1$ arrived at represents only the leading order approximation, which we denote as $\rho_1^{(0)}$. In this leading order, $\rho_2,\ldots,\rho_N$ are all zero. What we are interested in is the {\it next} order of approximation to $\rho_1$ as also to $\rho_2,\ldots,\rho_N$.

\vn One obtains $\rho_2,\ldots,\rho_N$ from the second,$\ldots$,$N$th equations in eq.~\eqref{eqA7} where, in the right hand side, we substitute $\zeta^{(0)}$ for $\zeta$, $\zeta^{(0)}$ being obtained from eq.~\eqref{eqA8} by replacing $\gamma_i$ with $\gamma_i^{(0)}=\frac{\rho_1^{(0)}}{\sqrt{N}}$ ($i=1,\ldots,N$). In other words, one has
\bs
\begin{align}
	\rho_i=\frac{K}{2}\lambda_i^{-1}\big(T^{\rm T}\zeta^{(0)}\big)_i,~(i=2,3,\ldots,N)\label{eqA13a}
\end{align}

\nn where
\begin{align}
	\zeta^{(0)}_i=\frac{2}{K}(\Delta-A sin (\delta-\phi_i-\frac{\rho_1^{(0)}}{\sqrt{N}})).\label{eqA13b}
\end{align}
\es

\nn Finally, denoting the correction to $\rho_1$ over $\rho_1^{(0)}$ by $\eta$,
\begin{align}
	\rho_1=\rho_1^{(0)}+\eta,\label{eqA13N}
\end{align}

\nn one determines $\eta$ by referring to the consistency condition~\eqref{eqA4}, now written with the correction terms included in $\gamma_i=(T\rho)_i$, i.e, with $\rho$ written as $\{\rho_1^{(0)}+\eta,\rho_2,\ldots,\rho_N\}^{\rm T}$, where $\rho_2,\ldots,\rho_N$ are determined from eq.~\eqref{eqA13a} as above. Having obtained now a consistent approximation for all components of $\rho$, one can finally work out $\gamma=T\rho$, and then the entrained solution~\eqref{eqA2}. 

\vn It may be noted that the effect of the periodic driving on a synchronized solution is two-fold, involving, first, the phase shifts $\gamma_i(i=1,2,\ldots,N)$ for the oscillators modifying the time-independent phases $\phi_i$ and, secondly, the change in the frequency from the free-running frequency $\omega$ to the driving frequency $\Omega$. Of these the phase shifts $\gamma_i$ are made up of a dominant term $\gamma_i^{(0)}$ which is the {\it same} for all the oscillators, along with correction terms determined by $\eta$ and $\rho_2,\ldots,\rho_N$ as given by eq.~\eqref{eqA13a}. However, a uniform phase shift for all the oscillators does not imply any change in the {\it orbit} of the system of $N$ oscillators in its phase space (a $N$-torus) since the orbit already represents a one-parameter family of solutions corresponding to various values of a uniform phase shift $\Phi$. Thus, the relevant part of the effect of the driving is contained in the correction terms mentioned above, apart from the shift in the frequency.

\subsection{Frequency window for a given driving amplitude}\label{window}

\nn An immediate consequence of the consistency condition~\eqref{eqA4} is that the detuning $\Delta$ and the driving amplitude $A$ have to satisfy  
\begin{align}
	\left|\Delta\right|<A,\label{eqA14}
\end{align}

\nn for an entrained solution (eq.~\eqref{eqA2}) to exist. In other words, for a given value of the driving amplitude $A$, the detuning $\Delta$ has to be within the window $-A<\Delta<A$. This condition requiring the detuning to be within a certain window depending on the driving amplitude is a common feature in experiments and models on entrainment (see, e.g.,~\cite{window}).

\begin{figure}[!h]
	\centering
		\includegraphics[width=0.8\textwidth]{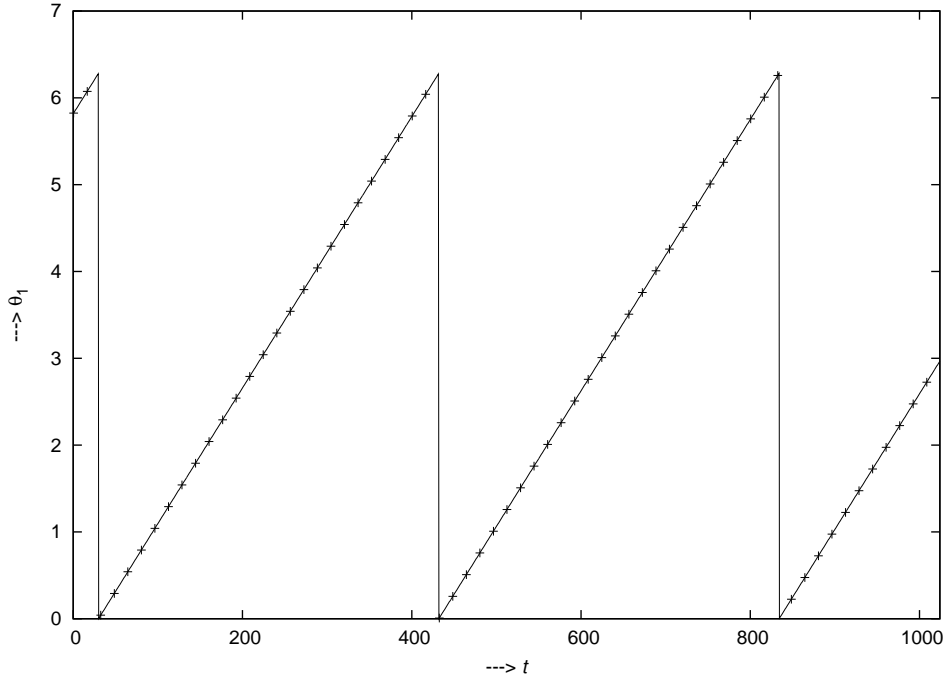}
		\caption{Depicting a stable entrained solution obtained by perturbation from a free-running rrr(0) one; $K=10.8066978, \delta=0, A=0.03125, \Delta=0.015625$; the curve frawn with a solid line corresponds to the solution obtained by the perturbation approach explained in sec.~\ref{entrain}, while the crosses are points generated by numerical integration; the agreement between the two is evident.}
	\label{kura6}
\end{figure}

\vn Fig.~\ref{kura6} depicts an entrained solution obtained by the above perturbative approach by starting from a free-running stable rrr(0) solution with $N=3$. Referring to sec.~\ref{entstab} below, one finds that this is a stable entrained solution for the system under consideration. One observes from the figure that the perturbation approach works excellently when compared with the time evolution obtained by a numerical integration of the system under consideration.

\subsection{Entrainment of intermittent chaos}\label{chao-entrain}

\begin{figure}[!h]
	\centering
		\includegraphics[width=0.8\textwidth]{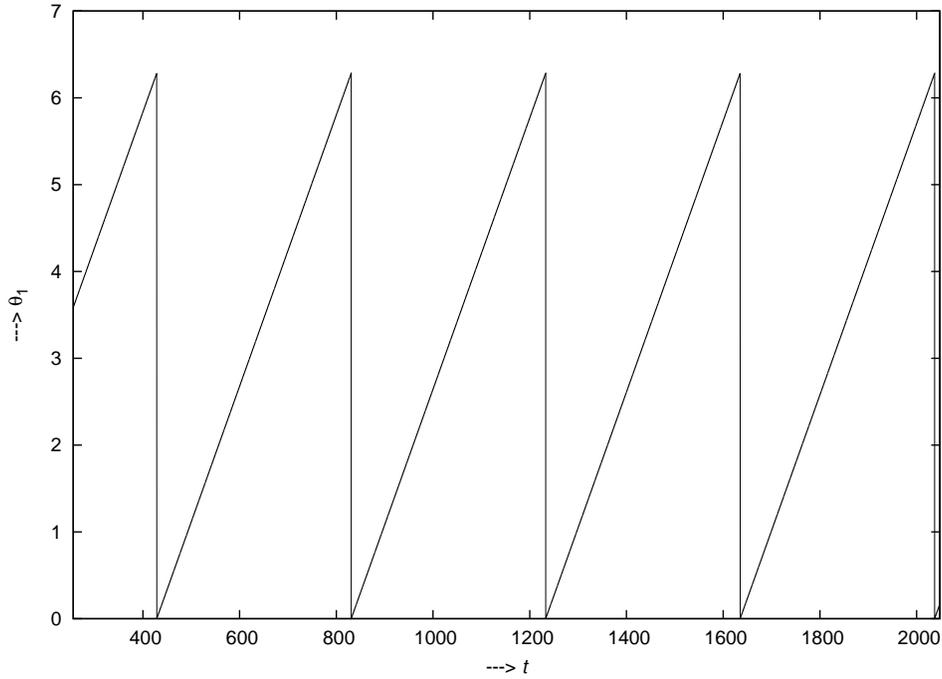}
		\caption{Entrainment of intermittent chaos; $K=K_{\rm c}-\frac{1}{256}$, $\delta, A$, and $\Delta$ as in fig.~\ref{kura6}, where $\Delta$ is now to be interpreted as the difference between $\Omega$ and the mean of the oscillator frequencies, rather than a free-running synchronized frequency; an initial condition is chosen close to $\theta_i=\phi_i$ mentioned in the first paragraph of sec.~\ref{chaos}, by numerically solving for $\gamma_i~(i=1,2,3)$ in eq.~\eqref{eqA3}.}
	\label{kura7}
\end{figure}

\vn Fig.~\ref{kura7}, on the other hand, depicts the phenomenon of {\it entrainment of intermittent chaos} where a periodic driving results in an entrained solution for the system (with $N=3$) for parameter values corresponding to intermittent chaos ($K<K_{\rm c}$). The driving frequency $\Omega$ for the existance of such an entrained solution has to satisfy once again the condition~\eqref{eqA14} where now $\omega(=\Omega-\Delta)$ is to be interpreted as the mean of the oscillator frequencies rather than the free-running frequency of an entrained solution.

\begin{figure}[!h]
	\centering
		\includegraphics[width=0.8\textwidth]{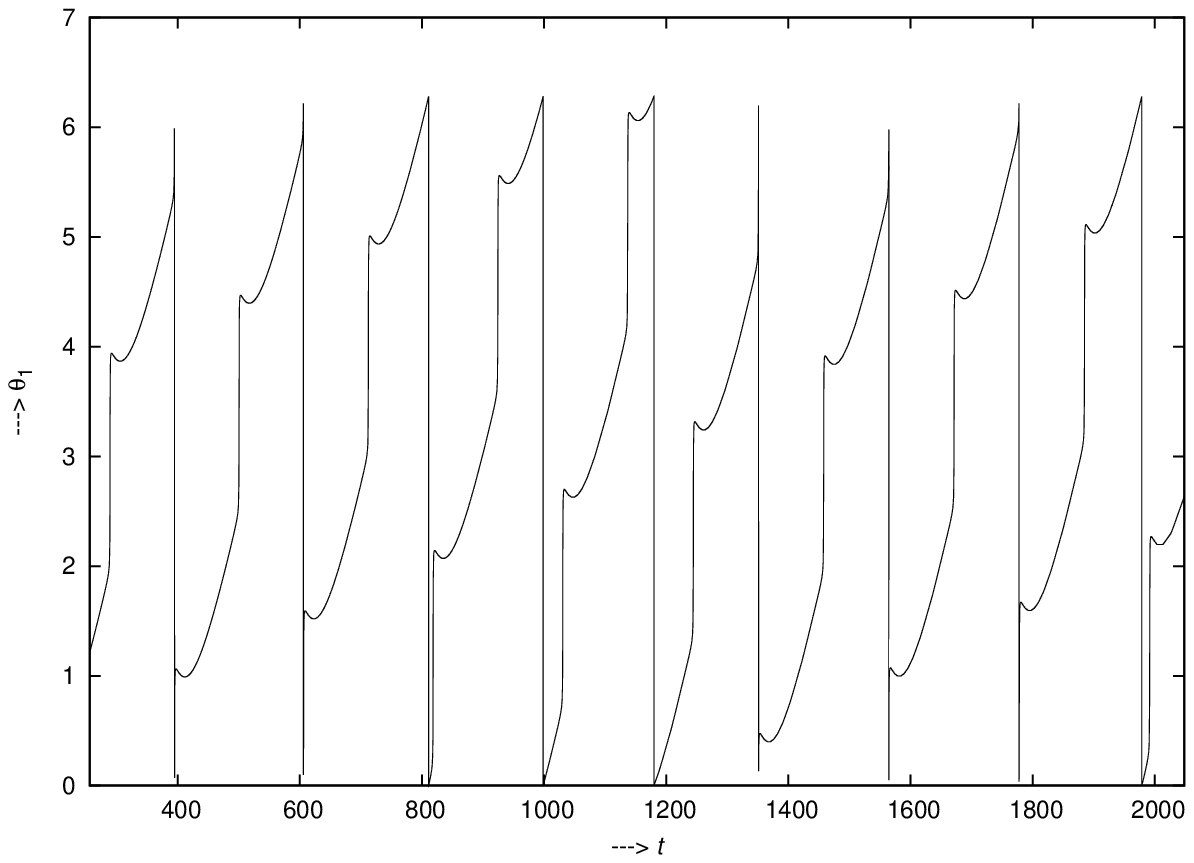}
		\caption{Irregular time variation of one of the three phase angles in the case of periodic driving on intermittent chaos ($K$ slightly less than $K_{\rm c}$; we choose $K=K_{\rm c}-\frac{1}{256}$) where the detuning is not small enough to satisfy the condition given by~\eqref{eqA14} (as in fig.~\ref{kura7}, $\omega$ is to be interpreted as the mean of the oscillator frequencies rather than the frequency of a free-running synchronized solution); we choose $\Delta=A+\frac{1}{1024}$ for the purpose of illustration.}
	\label{kura-extra}
\end{figure}

\vn By contrast, fig.~\ref{kura-extra} depicts a situation where the amplitude and frequency of the driving term violates the condition~\eqref{eqA14}, with $K$ slightly less than $K_{\rm c}$, in which one finds that entrainment does not take place, resulting instead in an irregular variation of the phase angles.

\subsection{Entrained solutions for large $A$}\label{largeA}

\nn While we have worked out the entrained solutions for small values of the driving amplitude (and, hence, of the detuning), there evidently exists entrained solutions for larger values of $A$ as well. However, the solutions for large $A$ do not necessarily have any correspondence with the free-running synchronized solutions obtained earlier. Thus, for $A>>K$, one has, as a first approximation, the entrained solution
\bs
\begin{align}
	\theta_i=\Omega t+\phi_i~(i=1,2,\ldots,N),\label{eqA15a}
\end{align}

\nn where the time-independent phases $\phi_i$, which now have no reference to the free-running phases, are given by
\begin{align}
	\phi_i=\delta-[sin^{-1}]\big(\frac{\Omega-\omega_i}{A}\big).\label{eqA15b}
\end{align}
\es

\nn Here the function $[sin^{-1}]$ differs from the principal value of the inverse sine in that it may correspond to either of the two branches $\alpha$ and $\beta$ introduced in sec.~\ref{synch}, i.e., to an angle in either of the two ranges $[0,\frac{\pi}{2})$ and $[\frac{\pi}{2},\frac{3\pi}{2})$.

\nn The condition for the existence of such an entrained solution is now (see eq.~\eqref{eq4b})
\begin{align}
	\left|\Omega-\omega_i\right|=\left|\Delta_i\right|<A~(i=1,2,\ldots,N),\label{eqA16}
\end{align}

\nn which, at the same time, implies condition~\eqref{eqA14}.

\vn Evidently, the above solution is stable if all the $\phi_i$'s are determined from eq.~\eqref{eqA15b} by choosing the branch $\alpha$ for $[sin^{-1}]$.

\vn For small values of $\frac{K}{A}$, entrained solutions can be obtained perturbatively from the above solutions determined for $A\rightarrow\infty$. Numerically, one obtains a host of such entrained solutions even away from the perturbative regime $\frac{K}{A}<<1$. However, the problem of continuation to the regime $\frac{A}{K}<<1$ remains.

\subsection{Stability of the entrained solutions}\label{entstab}
	
\vn Having obtained an entrained solution for either $\frac{A}{K}<<1$ or $\frac{K}{A}<<1$, or even away from these perturbative regimes by numerical computation, one can address the question of {\it stability} of the solution, which turns out to be a trivial one since the linearized equations are of the form
\begin{align}
	{\dot{\xi}}_i = \sum_j {M'}_{ij}\xi_j.\label{eq-stab1}
\end{align}

\nn Here $\xi_i~(i=1,2,\ldots,N)$ denote the deviations from the phases $\phi_i$ of the entrained solution (eq.~\eqref{eqA15a}) under consideration, and the matrix $M'$ is given (for $N=3$, for instance,) by
\begin{align}
	M'=\frac{K}{2}\begin{pmatrix}-(q_1+q_2+\frac{2A}{K}cos~\phi_1') & q_2 & q_1\\q_2 & -(q_2+q_3+\frac{2A}{K}cos~\phi_2) & q_3\\q_1 & q_3 & -(q_3+q_1+\frac{2A}{K}cos~\phi_3')\end{pmatrix}.\label{eq-stab2}
\end{align}

\nn In this expression, $q_i$ stands for $cos(\phi_i-\phi_{i-1})$ and $\phi_i'$ for $(\phi_i-\delta)$ ($i=1,2,\ldots, N$). In the perturbative regime $\frac{A}{K}<<1$, one has to substitute for $\phi_i$ the perturbed phase angles $(\phi_i+\gamma_i)$ where, in the latter expression, $\phi_i$ stands for the phase angles of the free-running solution and $\gamma_i$ for the perturbations considered in sec.~\ref{entrain}.

\vn The fact that {\it the stability matrix $M'$ is a constant rather than a periodic one} reduces the question of stability to a simple determination of the eigenvalues of $M'$.

\begin{figure}[!h]
	\centering
		\includegraphics[width=0.8\textwidth]{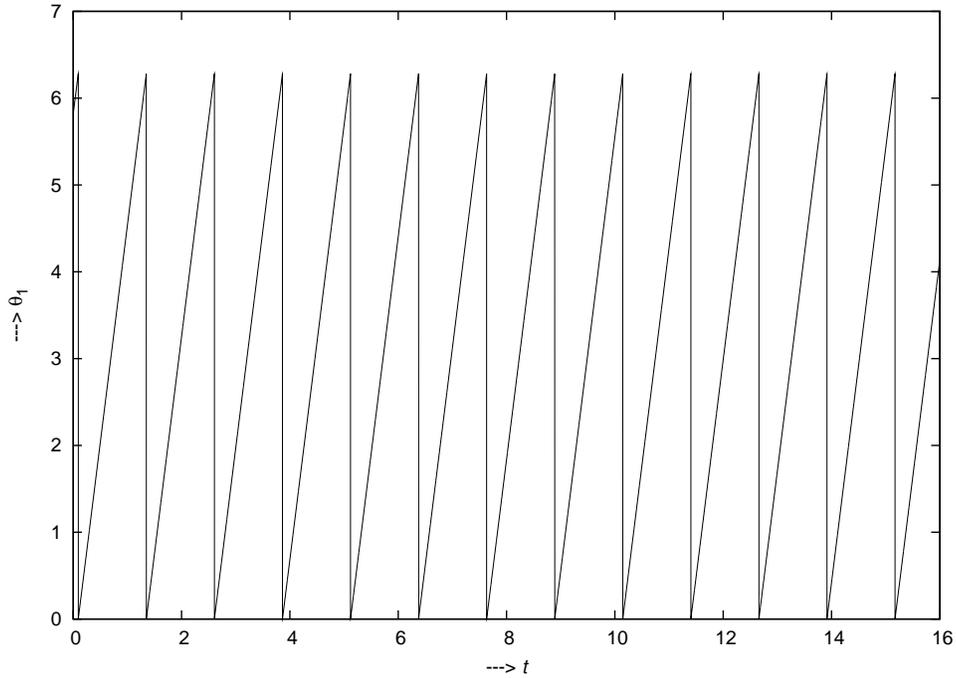}
		\caption{Depicting a stable entrained solution obtained numerically for $K=5,~A=10,~\delta=0,~\Omega=5$; these values correspond to the system being away from both the perturbative regimes $\frac{K}{A}<<1$ and $\frac{A}{K}<<1$.}
	\label{kura8}
\end{figure} 

\vn Fig.~\ref{kura8} depicts an entrained solution obtained numerically for $K=5$, $A=10$, $\Omega=5$, $\delta=0$, for which~\eqref{eqA16} is satisfied. The initial condition for the soilution was obtained by numerically solving eq.~\eqref{eqA3} for $\phi_i+\gamma_i~(i=1,2,3)$, starting from values given by eq.~\eqref{A15b} (where the principal value of $sin^{-1}$ was taken for all the three angles), and is a stable solution by virtue of all the eigenvalues of $M'$ (eq.~\eqref{eq-stab2}) being negative. It may be noted that the values of the parameters chosen correspond to the system being away from both the perturbative regimes mentioned above.


\begin{thebibliography}{}
\bibitem{chaos09}Rui Dilao, 'Antiphase and in-phase synchronization of nonlinear oscillators: The Huygens's clocks system',CHAOS {\bf 19}, 023118, 1-5, (2009).
\bibitem{strogatz}Steven H. Strogatz and Ian Stewart, 'Coupled Oscillators and Biological Synchronization', Scientific American, {\bf 251}, 102-109 (1993). 
\bibitem{buck} J. Buck and E. Buck, 'Synchronous fireflies', Scientific American, {\bf 234}, 74-85 (1976).
\bibitem{peskin} C. S. Peskin, 'Mathematical aspects of heart physiology', Courant Institute Of Mathematical sciences, New York University, N.Y., 268-278 (1975).
\bibitem{mirollo} R. E. Mirollo and S.H. Strogatz, 'Synchronization of Pulse-coupled Biological Oscillators', SIAM J. Appl. Math., {\bf 50}, 1645-1662 (1990).
\bibitem{ermentrout3} G. B. Ermentrout, 'The behaviour of rings of coupled oscillators', J. Math. Biol., {\bf 23}, 55-74 (1985).
\bibitem{kuramoto-pap} Y. Kuramoto and I. Nishikawa, 'Statistical macrodynamics of large dynamical systems: Case of a phase transition in oscillator communities', J. Stat. Phys., {\bf 49}, 569-605 (1987).
\bibitem{strogatz4} J. T. Ariaratnam and S. H. Strogatz, 'Phase Diagram for the Winfree Model of Coupled Nonlinear Oscillators', Phys. rev. Lett., {\bf 86}, 4278-4281 (2001). 
\bibitem{winfree} A.T.Winfree, 'Biological rhythms and the behaviour of populations of coupled oscillators',J. Theor. Biol., {\bf 16}, 15-42 (1967); A.T.Winfree, {\it The geometry of Biological Time}, Springer, N.Y., 1980.
\bibitem{kuramoto1} J. A. Acebron, L. L. Bonilla, C. J. P. Vicente, F. Ritort, and R. Spigler, 'The Kuramoto model: a simple paradigm for synchronization phenomena', Rev. Mod. Phys. {\bf 77}, 137-185 (2005).
\bibitem{sakaguchi} H. Sakaguchi, Cooperative Phenomena in Coupled Oscillator Systems under External Fields, Prog. Theor. Phys., {\bf 79}, 39-46 (1988).
\bibitem{ott5} T. M. Antonsen, R. T. Faghih, M. Girvan, E. Ott, and J. Platig, 'External periodic driving of large systems of globally coupled phase oscillators', Chaos, {\bf 18}, 037112, 1-10 (2008). 
\bibitem{circadian1} F. R. G. Cardoso, F. A. de O. Cruz, D. Silva, and C. M. Cortez, 'A simple model for circadian timing by Mammals', Brazilian Jour. Med. Biol. Research, {\bf 42}, 122-127 ()2009.
\bibitem{kuramoto5} Yu. Maistrenko, O. Popovych, O. Burylko, and P. A. Tass, 'Mechanism of Desynchronization in the fFinite-Dimensional Kuramoto Model', Phys. Rev. Lett., {\bf 93}, 084102, 1-4, (2004).
\bibitem{kuramoto8} O. V. Popovych, Yu. L. Maistrenko, and P. A. Tass, 'Phase chaos in coupled oscillators', Phys. Rev. E {\bf 71}, 065201(R), 1-4 (2005).
\bibitem{ermentrout1} L. Ren, and B. Ermentrout, 'Phase locking in chains of multiple-coupled oscillators', Physica D {\bf 143}, 56-73 (2000).
\bibitem{ermentrout2} L. Ren, and G. B. Ermentrout, 'Monotonicity of phase-locked solutions in chains and arrays of nearest-neighbour coupled oscillators', SIAM J. Math. Anal., {\bf 29}, 208-234 (1998).
\bibitem{ring-model} M. Golubitsky, I. Stewart, P. L. Buono, and J. Collins, 'A modular network for legged locomotion', Physica D, {\bf 115}, 56-78 (1998); C. Laing, 'Rotating waves in rings of coupled oscillators', Dyn. Stab. Syst., {\bf 13}, 305-18 (1998); L. Dussport, and J. Laheurte, 'Coupled oscillator array generating circular polarization', IEEE Microw. Guid. Wave Lett., {\bf 9}, 160-162 (1999).
\bibitem{rogge1} J. A. Rogge, and D. Aeyels, 'Stability of phase locking in a ring of unidirectionally coupled oscillators', J. Phys. A: Math. Gen., {\bf 37}, 11135-11148 (2004).
\bibitem{kuramoto6} J. Ochab, and P. F. Gora, 'Synchronization of Coupled Oscillators in a Local One-dimensional Kuramoto Model', Acta Physica Polonica B Proceedings Supplement, {\bf 3}, 453-462 (2010).
\bibitem{golub} G. H. Golub, and C. F. Van Loan, {\it Matrix Computations}, 3rd ed., The Johns Hopkins Univ. Press, Baltimore (1996).
\bibitem{mehta1} D. Mehta, and M. Kastner,  'Stationary point analysis of the one-dimensional lattice Landau gauge fixing functional, aka random phase XY Hamiltonian', arXiv:1010.5335v1 (2010). 
\bibitem{mehta2} D. Mehta, A. Sternbeck, L. von Smekal, and A. G. Williams, 'Lattice landau Gauge and Algenraic Geometry', arXiv:0912.0450v1 (2009). 
\bibitem{intermit} Y. Pomeau and P. Manville, 'Intermittent transition to turbulence in dissipative dynamical systems', Comm. Math. Phys., {\bf 74}, 189-197 (1980).
\bibitem{window} T. olde Scheper, D. Klinkenberg, C. Pennartz, and J. van Pelt, A mathematical Model for the Intracellular Circadian Rhythm Generator, Jour. Neuroscience, {\bf 19}, 40-47 (1999).

\end{thebibliography}
\end{document}